\documentclass{svproc}
\usepackage{url}

\usepackage{amsmath}
\usepackage{amssymb}
\usepackage{graphicx}

\usepackage{stackengine}
\usepackage{scalerel}

\def\hello{\stackon[0pt]{\ensuremath{{\bf v}\times{\bf b}}}{\hstretch{5.0}{\wedge}}\ensuremath{\!\!\!{}_{\,\,\bf k}}}

\begin{document}
\mainmatter              
\title{What makes a steady flow to favour kinematic magnetic field generation: A statistical analysis}

\titlerunning{Kinematic magnetic field generation: A statistical analysis}

\author{Francisco Stefano de Almeida\inst{1} \and Roman Chertovskih\inst{2} \and S\'ilvio Gama\inst{1}
\and Rui Gon\c{c}alves\inst{2}
}

\authorrunning{F.S. de Almeida {\it et al.}}

\institute{
Department of Mathematics, Faculty of Sciences, University of Porto. Rua do Campo Alegre s/n
4169-007 Porto, Portugal 
\email{smgama@fc.up.pt, francisco91stefano@gmail.com}
\and
Research Center for Systems and Technologies (SYSTEC), ARISE, Faculty of Engineering, University of Porto.  Rua 
Dr. Roberto Frias s/n, 4200-465 Porto, Portugal
\email{roman@fe.up.pt, rjpg@fe.up.pt}
}

\maketitle

\begin{abstract}

To advance our understanding of the magnetohydrodynamic (MHD) processes in liquid metals, in this paper we propose an approach combining the classical methods in the dynamo theory based on numerical simulations of the partial differential equations governing the evolution of the magnetic field with the statistical methods. In this study, we intend to answer the following ``optimization'' question: Can we find a statistical explanation what makes a flow to favour magnetic field generation in the linear regime (i.e. the kinematic dynamo is considered), where the Lorenz force is neglected? The flow is assumed to be steady and incompressible, and the magnetic field generation is governed by the magnetic induction equation. The behaviour of its solution is determined by the dominant (i.e. with the largest real part) eigenvalue of the magnetic induction operator. Considering an ensemble of 2193 randomly generated flows, we solved the kinematic dynamo problem and performed an attempt to find a correlation between the dominant eigenvalue and the standard quantities used in hydrodynamics -- vorticity and kinetic helicity. We have found that there is no visible relation between the property of the flow to be a kinematic dynamo and these quantities. This enables us to conclude that the problem requires a more elaborated approach to ``recognize'' if the flow is a dynamo or not; we plan to solve it using contemporary data-driven approach based on deep neural networks. 

\keywords{Magnetohydrodynamics, Kinematic Dynamo, Dominant Eigenvalue, Pseudospectral Methods.}

\end{abstract}

\section{Introduction}

Magnetic field protects the Earth from cosmic radiation, this being necessary for the existence of life. The long periods of magnetic field quasistationarity are interrupted by reversals \cite{J94}, when in a relatively short time the field polarity is changed to the opposite. If during a reversal the field vanishes, it stops shielding the planet, oxygen escapes from the atmosphere \cite{W14}, the ozone layer is destroyed \cite{M16} and the climate drastically changes \cite{V14}.

Paleomagnetic records reveal that low magnetic field periods usually coincide with warm climate periods \cite{W71}. Physical mechanisms responsible for the correlation of climatic and geomagnetic variations are not yet fully clear; for instance, the weakening of the geomagnetic field may cause an increase in cyclonic activity~\cite{B76} or the higher impact of cosmic rays on the cloud formation \cite{K11}. In view of such links, the evolution of the geomagnetic field is indispensable for a better understanding of the climate changes \cite{K20}.

According to \cite{V16}, we may at present experience the beginning of a reversal; thus, studying this phenomenon is an urgent scientific task of great practical importance. In local anomalies of the present time, the magnetic field is weaker, and thus protection from the radiation of humans at the ground level is weaker as well, causing more deaths from cancer. Some studies \cite{PC16} interpret the South Atlantic Anomaly as an indicator of an upcoming geomagnetic transition.

Magnetic fields of planets \cite{S03} owe their existence to the convection of electrically conducting media in their interiors. The dynamo theory \cite{R13,M19} addresses various phenomena of the dynamic origin of the terrestrial magnetic field (reversals, excursions, etc. \cite{G20,T23}). Convective dynamos are studied by solving numerically equations of magnetohydrodynamics \cite{G95,B08}. Even modern supercomputers are not powerful enough to resolve all scales in the interior of the Earth. Thus, simplified models must be considered, e.g. the kinematic dynamo problem, where the influence of the generated magnetic field on the flow by the Lorentz force is neglected. Many dynamo models are based on the mean-field electrodynamics \cite{K80,Z83}, where some empirical relations for the transport coefficients are postulated, however, the validity of this approach has been critically assessed, see, e.g., \cite{S07,C15,R18,AD19,AP19}.

The first results on geodynamo modelling were reported in \cite{G95}, where the dipolar structure of the geomagnetic field was reproduced. Since then numerous simulations of convective dynamos have been performed; results of the most recent numerical experiments are reviewed in \cite{W19}. However, the values of the parameters used differ in many orders of magnitude from the realistic ones. In order to understand such complex temporal dynamics, the main mechanisms responsible for magnetic field generation are crucial to understand, e.g. considering the simplest setup -- the kinematic dynamo problem \cite{S18}.

If a fluid flow is prescribed, the evolution of the magnetic field is governed by the magnetic induction equation \cite{M19} -- second-order parabolic partial differential equation. The flow, responsible for magnetic field generation, enters the equation as a parameter. If the flow is steady, the solution of the magnetic induction equation is determined by the real part of the dominant eigenvalue (possessing the maximal real part among all eigenvalues) of the magnetic induction operator. If the real part of the dominant eigenvalue is positive, then an initial seed magnetic field grows in time; if it is negative, the initial field decays to zero. In the former case, the flow is called to be a kinematic dynamo.

Physicists ask the following question: why does one flow of a conducting fluid generate a magnetic field better than the other? The mathematical answer to this question is that the real part of the dominant eigenvalue of the magnetic induction operator is larger for one flow and smaller for the other. Unfortunately, the physicists find this answer unsatisfactory. Sometimes this question can be answered for a particular family of flows (see, e.g., \cite{C17}), but such results are not universal. Our main goal is to find an alternative answer to this question. 

As an ultimate step of this project, we expect that methods based on data science, e.g. deep neural networks, can be successful in solving with reasonable precision the kinematic dynamo problem. If this is the case, then using recent methods of explainable learning, we will be able to extract features in the conducting fluid flows responsible for magnetic field generation. The goal is to outline a geometric description of flows that amplify dynamo production, similar to, for instance, what was done for two-dimensional incompressible isotropic flows, invariant by sixty-degree rotations, leading to negative eddy viscosities. In this case,  the focus lies on pinpointing regions characterized by densely packed streamlines, which locally resemble the Kolmogorov flow~\cite{GVF}.

We expect our project to be successful because there is a similarity between the kinematic dynamo problem and the image classification problem successfully solved by convolutional neural networks (CNNs) \cite{G16}. From the image classification point of view, the flow-generating magnetic field is an “image”, which can be “classified” to be more or less beneficial for the magnetic field generation. This analogy becomes clearer for the simplest flows, considered in this paper (dependent only on two spatial variables): a flow ${\bf v}=(v_1,v_2,v_3)$ on a uniform grid is a two-dimensional array (“image”), where each element (“pixel”) contains a component $v_i$ of the flow (“RGB intensity”). This convinces us that the proposed data-driven approach for the kinematic dynamo problem is promising and, hence, is worth studying. 

Before applying data-science methods, in the present paper, we presents results of a much simpler analysis, based on the statistical analysis. In other words, we test the idea if ``what makes a flow to be a strong/weak dyanmo'' can be characterised by a measure like a norm of vorticity or kinetic helicity, or can be detected visually. We synthesize 2193 sample steady flows and, for each flow,  solve the kinematic dynamo problem (these data will be also used as a dataset in future for the machine learning approach) on a high performance computer. Then we explore if the kinematic dynamo problem can be related statistically to a certain characteristic if the generation fluid flow. We also check if the statistical approach will let us extract features in the conducting fluid flow responsible for magnetic field generation.

The paper is organised as follows. In the next section we formulate the kinematic dynamo problem and discuss its numerical solution. In Section~\ref{sec:vels} we present how the sample flows were constructed, to be used further in the statistical analysis presented in Section~\ref{sec:stat}. Our conclusions and plans how to continue this line of research are given in the last section. 

\section{Kinematic dynamo problem \label{sec:kinem}}

\subsection{Statement of the problem}

All fields (flow and magnetic field) are assumed to be periodic in all spatial directions, $x_1$, $x_2$ and $x_3$, with the same period of $2\pi.$ We denote ${\bf x}=(x_1, x_2, x_3)$. A small random magnetic field is given at the initial time. The magnetic field is assumed to be solenoidal and fluid flow is incompressible, the spatial means of the flow and magnetic field are assumed to vanish.

To create the dataset to be analysed, we use steady 2.5D flows (i.e. depending only on two spatial variables) of an incompressible fluid, 
$${\bf v}=\left(v_1(x_1, x_2),\, v_2(x_1, x_2),\, v_3(x_1, x_2)\right).$$
These flows are synthesised randomly in the Fourier space (see details in the next section). We note that we do not consider two-dimensional flows, because they do not generate magnetic fields by the Zeldovich antidynamo theorem \cite{Z83}.

We consider the kinematic dynamo problem, i.e. evolution of magnetic field governed by the magnetic induction equation: 
\begin{equation}
\label{eq:kinem}
\frac{\partial {\bf b}}{\partial t}=\mathcal{L}_{\bf v}\left({\bf b}\right), 
\end{equation}
where
\begin{equation}
\label{eq:op}
\mathcal{L}_{\bf v}\left({\bf b}\right) = \eta \nabla^2 {\bf b} + \nabla \times \left( {\bf v} \times {\bf b} \right)    
\end{equation}
is the magnetic induction operator, the flow ${\bf v}$ is steady, and $\eta$ is (constant) molecular magnetic diffusivity. 

Magnetic field is solenoidal
\begin{equation}
    \label{eq:div0}
    \nabla \cdot {\bf b}=0.
\end{equation}
If the condition (\ref{eq:div0}) is satisfied at the initial time, $t=0$, then, by virtue of (\ref{eq:kinem}) it is satisfied at any time $t$.

Let $\langle \cdot \rangle$ denote the (spatial) mean over the periodicity cell, $[0, 2\pi]^3$, of a scalar or vector field: 
$$
\langle {\bf f} \rangle = \frac{1}{(2\pi)^3} \int_{[0,2\pi]^3} {\bf f}({\bf x})\ {\rm d}{\bf x}.
$$

We assume the magnetic field to be zero-mean, i.e. $\langle {\bf b} \rangle$=0. If this property of the magnetic field is satisfied initially, it is satisfied at any time by virtue of the governing equation (\ref{eq:kinem}). In what follows we will use the norms:
\begin{equation}
\label{eq:norms}
\|{\bf f}||^2_2=\langle {\bf f}\cdot{\bf f} \rangle, \quad 
\|{\bf f}\|^2_\infty=\max_{{\bf x}\in[0,2\pi]^3} |{\bf f}\cdot{\bf f}|\,.
\end{equation}

Equation (\ref{eq:kinem}) is a linear in ${\bf b}$ parabolic partial differential equation of second order. Evolution of a weak initial magnetic field is determined by the {\it dominant eigenvalue}, $\lambda_d$ (whose real part is the {\it growth rate}) of the magnetic induction operator, i.e. the solution to the eigenvalue problem
\begin{equation}
    \label{eq:eigen}
    \mathcal{L}_{\bf v}\left({\bf b}\right)=\lambda_d\, {\bf b}\,, 
\end{equation}
where the eigenvalue has the maximal real part among all the eigenvalues $\lambda$ and the eigenfunctions ${\bf b}(\bf x)$ satisfy the solenoidality condition (\ref{eq:div0}). Dominant eigenvalue is also known as spectral abscissa \cite{V09}. 

\subsection{Numerical solution}
To compute dominant eigenvalue of the eigenvalue problem (\ref{eq:eigen}) we use the pseudospectral methods \cite{P02} and a variant of the Krylov subspace method \cite{Z93}.

Computations of the linear operator are performed for the magnetic field represented by its Fourier series: 
\begin{equation}
\label{eq:Four}
{\bf b}({\bf x})=
\sum_{k_1=-N/2+1}^{N/2-1}
\sum_{k_2=-N/2+1}^{N/2-1}
\sum_{k_3=-N/2+1}^{N/2-1}
\widehat{\bf b}_{\bf k} \,\, {\rm e}^{i {\bf k}\cdot {\bf x}},
\end{equation}
where the wave vector is denoted by ${\bf k}=(k_1, k_2, k_3)$. Only a half (for $k_1\ge0$) of the Fourier coefficients of magnetic field, $\widehat{\bf b}_{\bf k}$, in (\ref{eq:Four}), are stored in memory due to mutual complex conjugacy $$\widehat{\bf b}_{\bf k}=\widehat{\bf b}_{-\bf k}^*\,.$$ 

The solenoidality condition (\ref{eq:div0}) in the Fourier space satisfies $${\bf k}\cdot\widehat{\bf b}_{\bf k}=0\,.$$ 

The flow field ${\bf v}({\bf x})$ is assumed to be given in the physical space. All derivatives in (\ref{eq:op}) are computed in the spectral space, the cross product of the flows and magnetic field is computed in the physical space, forward and backward fast Fourier transforms (FFT) are used to switch between the spaces.

In particular, in the Fourier space the magnetic induction operator is represented by the Fourier series of the form (\ref{eq:Four}) with the Fourier coefficient (corresponding to the wave vector ${\bf k}$):
\begin{equation}
    -|{\bf k}|^2 \widehat{\bf b}_{\bf k} + i\sum_{n=1}^3 k_n\, {\bf e}_n \times \hello\,,
\end{equation}
where
\hello\,\,  stand for the Fourier coefficient of the corresponding cross product.

The code implementing this algorithm has been  proved useful in our previous works on this subject \cite{R18,AD19,AP19,Z20,RJES2}, it is written in FORTRAN 2003. Forward and backward FFTs implemented in the FFTW library \cite{FFTW} were used. Most of the computations were performed in parallel on several 18-core 3 GHz Intel Xeon Gold 6154 CPU (each core was performing a separate run), using Intel FORTRAN compiler, version 2021.6.0. 

\section{Construction of sample fluid flows\label{sec:vels}}

In order to investigate magnetic field generation by steady flows, we have constructed sample steady, solenoidal, zero-mean, $2\pi$-periodic (in each spatial coordinate) fields ${\bf v}=(v_1, v_2, v_3)$. All the fields are synthesized in the Fourier space using the discrete Fourier series analogous to (\ref{eq:Four}).  

The following procedure has been employed:
\begin{itemize}
\item White-noise three-dimensional pseudo-random complex vectors $\widehat{\bf v}_{\bf k}$ are seeded in the Fourier space for $k_1\ge0$ and $k_3=0$ of wave vectors ${\bf k}$. We use the random generator subroutine ${\tt random\_number}$ in order to generate real and complex parts of the Fourier harmonics with a uniform distribution in $(-0.5,\,0.5)$.   
\item The condition $\widehat{\bf v}_{\bf k}=\widehat{\bf v}_{-\bf k}^*$ is used
to initialize the remaining Fourier coefficients (so that the resultant field is real) and is imposed on the coefficients for $k_1=0$.
\item The mean part $\langle {\bf v} \rangle$ is removed (setting $\widehat{\bf v}_{\bf 0}=0$) and the solenoidality condition (orthogonality of the Fourier coefficient to the respective wave vector, ${\bf v}_{\bf k}\cdot {\bf k}=0$) is enforced by removing
the gradient part $({\bf v}_{\bf k}\cdot {\bf k}){\bf k}/|{\bf k}|^2$.
\item Each Fourier coefficient $\widehat{\bf v}_{\bf k}$ is divided by $2.5^{|{\bf k}|}$ (resulting in the exponential energy spectrum decay). 
\item Using the Parseval's identity, the field is normalized so that its r.m.s.~value, $\sqrt{\langle {\bf v}\cdot{\bf v} \rangle}$, is equal to 1.
\item Finally, the constructed flow in the Fourier space is computed in the physical space, where it is stored for the computations of the kinematic dynamo problem.
\end{itemize}
All flows are generated by this procedure; different flows are generated changing the seed for the pseudo-random number generator (calling {\tt random\_seed} subroutine) before implementing the first step of the algorithm above. In total, 2193 flows were generated by this procedure. With the considered resolution of $N=32$ (i.e. $|k_i|\le 15,\ i=1,2,3$), the energy spectra of the flows decrease by 9 orders of magnitude.  

\section{Statistical analysis \label{sec:stat}}

\begin{figure}
    \centering
    \includegraphics[width = \textwidth]{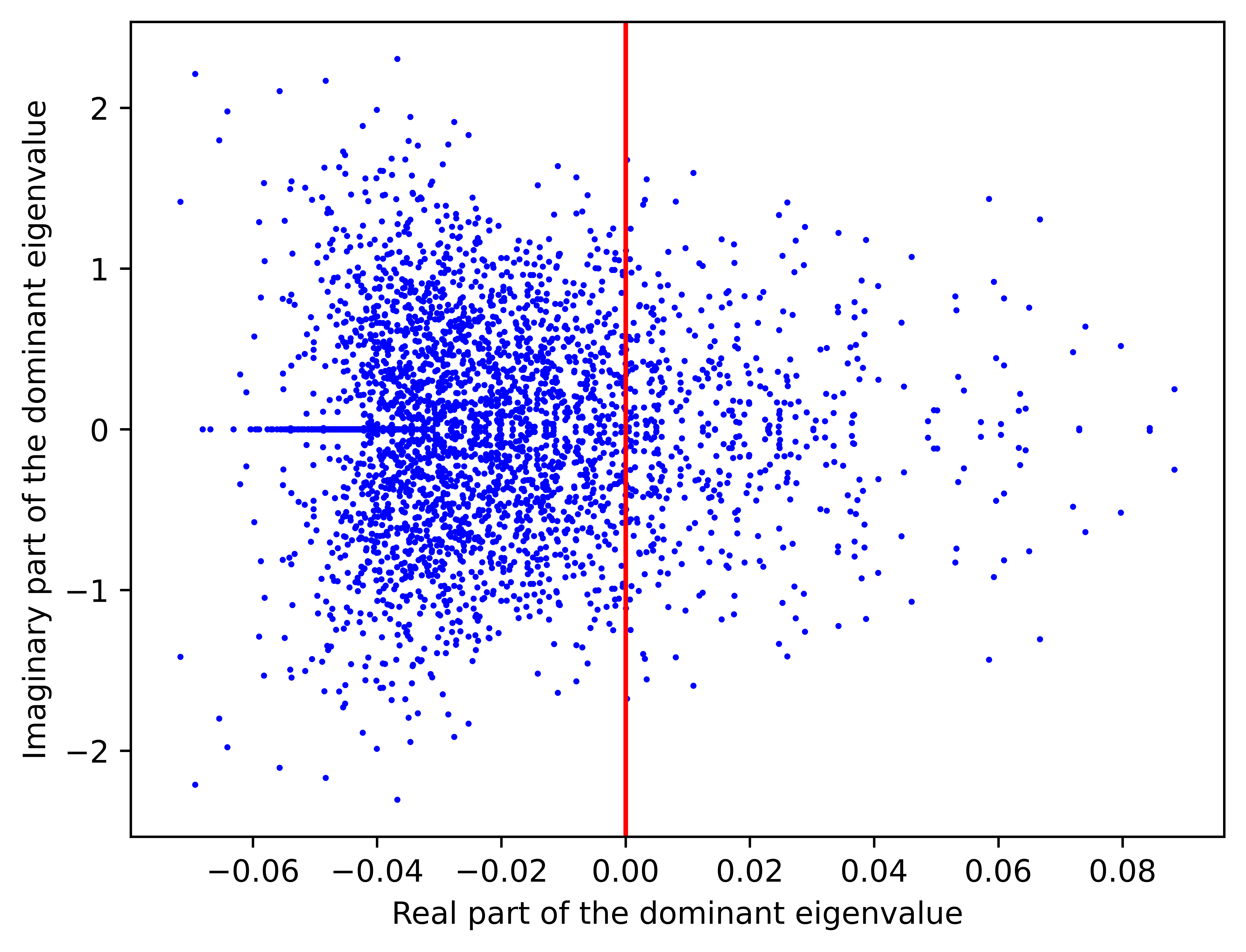}
    \caption{Dominant eigenvalues, $\lambda_d\,,$ of the magnetic induction operator. The vertical red line separates non-dynamos (left half-plane) from dynamos (right half-plane).}
    \label{fig:dom}
\end{figure}

\begin{figure}
    \centering
    \includegraphics[width = \textwidth]{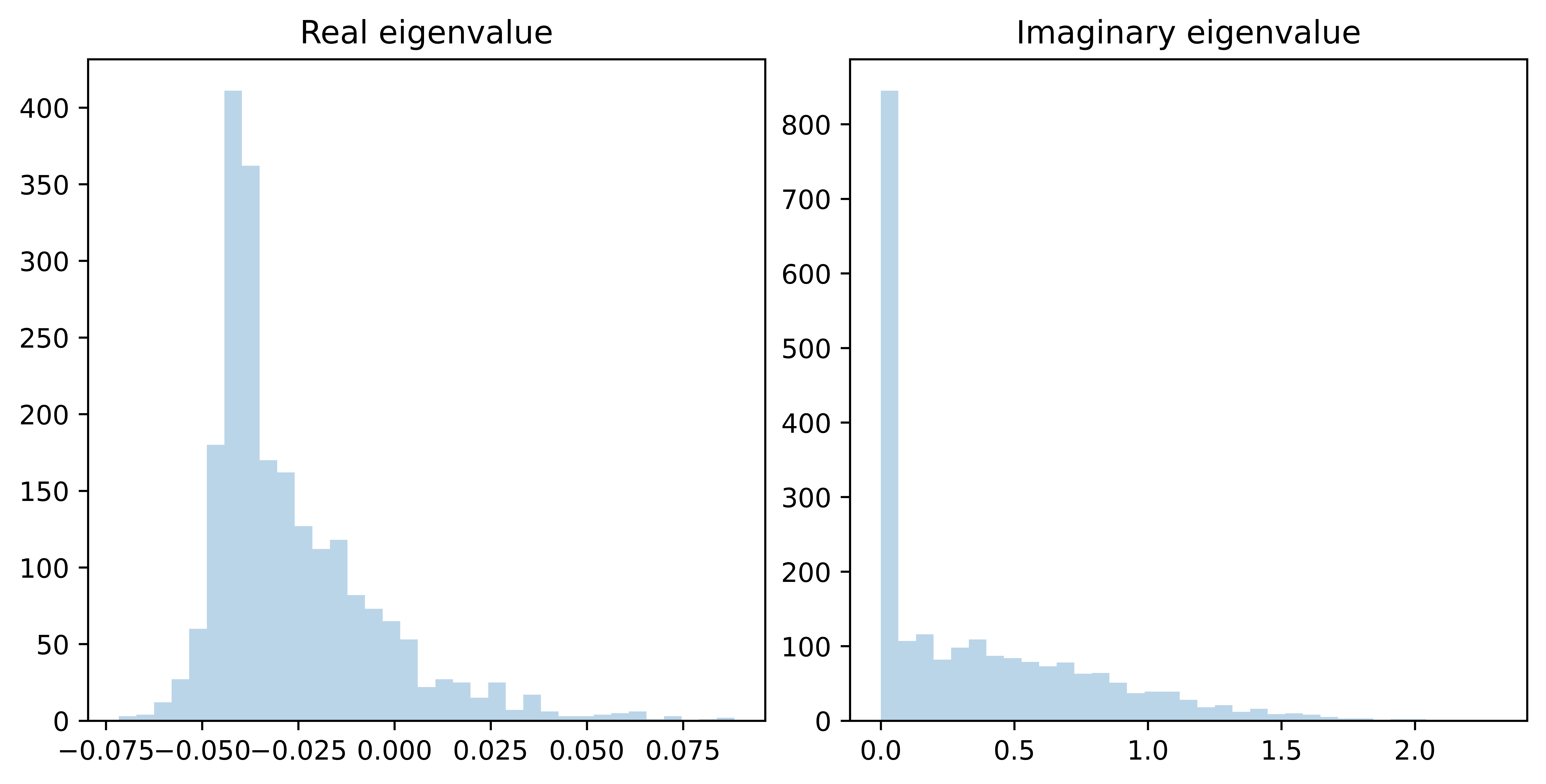}
    \caption{Histogram of the real and imaginary parts of the dominant eigenvalues of the magnetic induction operator. For complex conjugated pairs only the eigenvalue with non-negative imaginary part is used.}
    \label{fig:histo}
\end{figure}

We solved the kinematic eigenvalue problem for $\eta=0.03$ and each of the 2193 randomly generated flows described in the previous section. In Fig.~\ref{fig:dom}, the dominant eigenvalues of the magnetic induction operator for each of the 2193 velocity base flows are depicted. The graph exhibits symmetry about the horizontal axis due to the Hermitian symmetry. The histograms of real and imaginary parts are shown in Fig.~\ref{fig:histo}. 

In computations, we used numerical resolution of $N=32$ harmonics in every spatial direction. The energy spectra of all eigenfunctions decrease at least by 4 orders of magnitude, confirming that reasonable numerical resolution was employed. Although from the conservative viewpoint this is the minimum acceptable fall off, the growth rates differ from those computed with the doubled resolution (we did this test for some ten sample flows) only in the fifth digit.

The computed dominant eigenvalues are 730 real (33\%), all other 1463 dominant eigenvalues are complex conjugated pairs.

The hydrodynamic quantities we use to test for correlation with the magnetic field growth rate are: the norms (\ref{eq:norms}) of the vorticity and kinetic helicity and the mean kinetic helicity. Pearson correlation coefficient between the real part of the dominant eigenvalue and the hydrodynamic quantitites are presented in 
Table~\ref{tab:my_label}. These results show no significant correlation between the property of the flow to generate magnetic field and the considered hydrodynamical quantities. 

\begin{table}[t]
    \centering
    \begin{tabular}{|c|c|c|c|c|}\hline
       $\|\nabla \times {\bf v}\|_2$  & $\|\nabla \times {\bf v}\|_\infty$ & 
       $\langle{\bf v} \cdot\left( \nabla \times {\bf v}\right)\rangle$ & 
       $\|{\bf v} \cdot\left( \nabla \times {\bf v}\right)\|_2$ &   $\|{\bf v} \cdot\left( \nabla \times {\bf v}\right)\|_\infty$\\\hline
      0.0786 & 0.0808 & -0.0166 & 0.1831 & 0.1225 \\\hline 
    \end{tabular}
    \caption{Pearson correlation coefficients between the real part of the dominant eigenvalues and the correspondent global hydrodynamics quantities listed in the first line.}
    \label{tab:my_label}
\end{table}

\begin{figure}
    \centering
    \includegraphics[width = \textwidth]{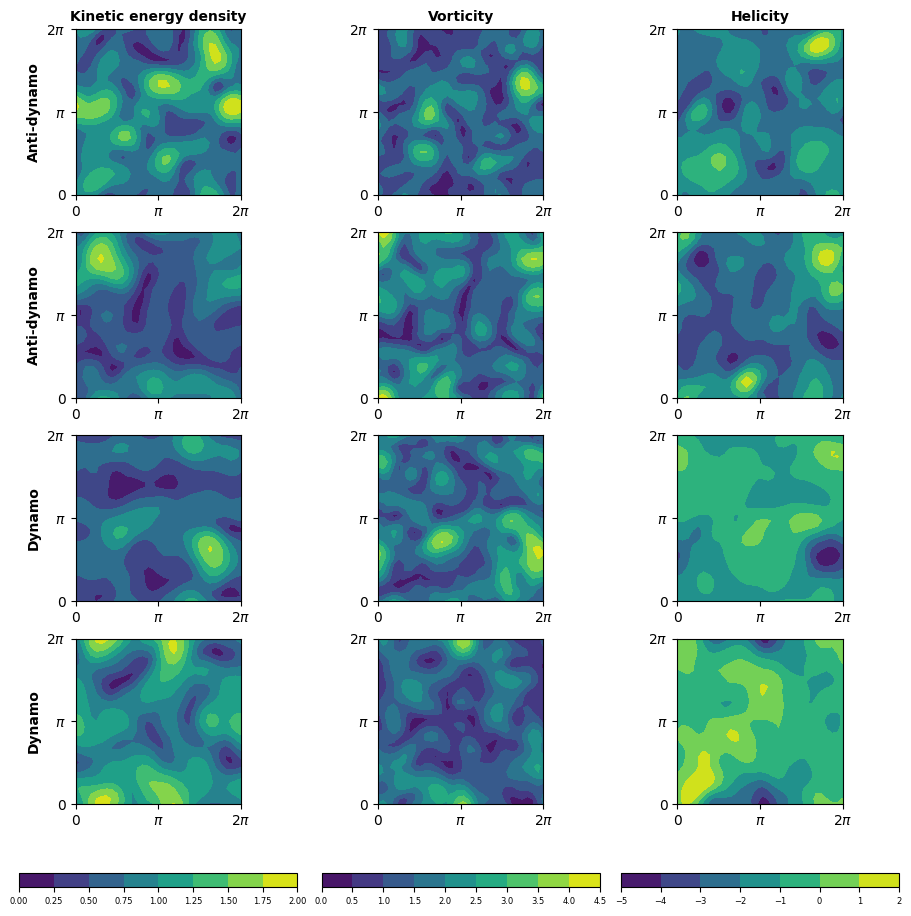}
    \caption{Contour plots of the kinetic energy density (first column), vorticity (second column) and kinetic helicity (third column) for the flows, which are  least (first two lines) and most (last two lines) favourable to magnetic field generation. The growth rates of the magnetic fields are -0.072 (first line), -0.069 (second), 0.084 (third) and 0.088 (fourth).}
    \label{fig:cont}
\end{figure}

We also considered two most favorable to magnetic field generation fields (corresponding to the growth rates 0.084 and 0.088) and two least favourable (growth rates -0.072 and -0.069). In order to find structures, which could distinguish the dynamos from no-dynamos, we plotted for these four flows the kinetic energy density ${\bf v}\cdot{\bf v}$, magnitude of vorticity, $|\nabla \times{\bf v}|$ and kinetic helicity density in the plane $(x_1,\, x_2)$, see Fig.~\ref{fig:cont}. Again, no visible differences in the flows demonstrating opposite properties in magnetic fields generation are detected. 

\section{Conclusions and future work}

We generated 2193 sample steady flows and for $\eta=0.03$ computed the dominant eigenvalues, determining the growth or decay of initially small magnetic field. We have found that for these flows there is no strong correlation between the norms of the standard hydrodynamic quantities and the increment of the magnetic field. This certifies that the kinematic dynamo problem is challenging and there is no simple characteristics distinguishing a field to be a dynamo is found by now. We believe that more sophisticated contemporary machine learning methods can be successful in solving this problem. 

The same data set presented here (inputs: flows; outputs: dominant eigenvalue) will also be used at the next step of this line of research, where we plan to apply CNNs to learn from the data and predict for a given flow its ability to generate magnetic field. First, we will find out if this problem can be solved by CNN with reasonable precision. Second, applying the Grad-CAM technique~\cite{S20}, producing “visual explanations” for decisions, to identify and analyse the regions of the flow responsible for the decision taken by CNN. Analysing the regions responsible for a flow to a dynamo, we will try to find a physical quantity (e.g. vorticity, flow helicity, kinetic helicity, etc.), which is high in those regions. This will let us conclude what makes a flow to be more favourable for magnetic field generation. 

We plan that our results not only advance the dynamo theory itself, but also provide information for dynamo experiments which flows are more beneficial for magnetic field generation, i.e. which flows provide faster growth of a seed magnetic field (also, in our experience, flows, which are better kinematic dynamos, demonstrate stronger magnetic fields in the saturated non-linear regime). If our approach will prove useful, we will extend it in future to the problems, also of interest in geophysics, related to evolution of large-scale magnetic fields, describing polarity reversals of the geomagnetic field, and to engineering magnetohydrodynamical problems on cooling of nuclear/fusion reactors by liquid metals.

\section*{Acknowledgments}
RC acknowledges the financial support by the FCT  doi:10.54499/CEECINST/ 00010/2021/ CP1770/CT0006. This work was also supported by multiple funding sources including the: Base funding (UIDB/00147/2020) and Programmatic funding (UIDP/00147/2020) of the SYSTEC -- Center for Systems and Technologies; ARISE - Associate Laboratory for Advanced Production and Intelligent Systems (LA/P/0112/2020) and the RELIABLE project (PTDC/EEI-AUT/3522/ 2020) funded by national funds through the FCT/MCTES (PIDDAC). A part of the computations was carried out on the OBLIVION Supercomputer (Évora University) under FCT computational project 2022.15706.CPCA and on the Google Cloud Platform, project ref. CPCA-IAC/AV/594467/2023. SG and FSA thanks the support by CMUP, a member of LASI, financed by national funds through FCT – Fundação para a Ciência e a Tecnologia, I.P., under the project with reference UIDB/00144/2020.

\bibliographystyle{splncs}

\end{document}